# Non-intrinsic origin of the Colossal Dielectric Constants in $CaCu_3Ti_4O_{12}$

P. Lunkenheimer,[1] R. Fichtl,[1] S.G. Ebbinghaus,[2] and A. Loidl[1]

[1]*Experimentalphysik V, Center for Electronic Correlations and Magnetism, Universität Augsburg, D-86135 Augsburg, Germany*
[2]*Festkörperchemie, Institut für Physik, Universität Augsburg, D-86135 Augsburg, Germany*

The dielectric properties of $CaCu_3Ti_4O_{12}$, a material showing colossal values of the dielectric constant, were investigated in a broad temperature and frequency range extending up to 1.3 GHz. A detailed equivalent circuit analysis of the results and two crucial experiments, employing different types of contacts and varying sample thickness, provide clear evidence that the apparently high values of the dielectric constant in $CaCu_3Ti_4O_{12}$ are non-intrinsic and due to electrode polarization effects. The intrinsic properties of $CaCu_3Ti_4O_{12}$ are characterized by charge transport via hopping of localized charge carriers and a relatively high dielectric constant of the order of 100.

Recent reports on the observation of colossal dielectric constants (CDCs) in $CaCu_3Ti_4O_{12}$ (CCTO) reaching values up to $10^5$ [1,2,3] have generated considerable interest in this material and related compounds [4,5,6,7,8,9,10,11,12,13,14]. The stunning observation in CCTO was a high and almost temperature-independent dielectric constant $e'$ at elevated temperatures and a steep decrease of almost three orders of magnitude at low temperatures. The step-like decrease of $e'$ as function of temperature and a concomitant peak in the dielectric loss $e''$, strongly depend on the measuring frequency and roughly follow an Arrhenius behavior. This relaxational behavior was ascribed to the slowing down of highly polarizable relaxational modes [2] or to the slowing down of dipolar fluctuations in nanosize domains [3]. Similar effects were obtained in thin films of CCTO making this system a good candidate for many applications [7].

However, shortly after the reports of the CDCs, their intrinsic nature has been questioned and arguments have been put forth that extrinsic effects as contributions from the electrode/sample interface, from grain boundaries in polycrystalline materials, or from twin boundaries in single crystals may be the sources of the giant dielectric constant [4,10,12,14,15]. And indeed, as is known since decades [16,17], the above outlined characteristics of the dielectric permittivity exactly corresponds to what is expected in case of interfacial polarization. Similar behavior, termed Maxwell-Wagner relaxation, has been observed in numerous materials (see, e.g., [15]). It was pointed out that, even if the CDCs in CCTO are not intrinsic, this material could be a possible candidate for commercial application as internal barrier layer capacitor (IBLC), i.e. for the case of internal barriers at grain boundaries, leading to the observed high values of $e'$ [5]. Thus, clearly there is an urgent need for clarification of the true origin of the observed CDCs, which in case of an intrinsic nature would have important theoretical implications and in case of an IBLC scenario still would be of high technical relevance.

As has been pointed out in [15], intrinsic (or IBLC) and electrode effects can well be separated using different contacts and sample geometries. Consequently, Ramirez *et al.* [8] remeasured CCTO using a different type of contact preparation. Despite the fact that they observed significantly lower absolute values of $e'$ compared to their earlier work [2], they provided arguments in favor of an intrinsic effect and explained the CDC in terms of braced lattices with defects. However, as will be discussed below, these experiments are not a final proof. The aim of the present work is to finally clarify, whether or not the detected CDCs in CCTO are due to external contact contributions. For this purpose we have measured the complex dielectric permittivity and conductivity of CCTO over a broader frequency range than in all previous reports, extending up to 1.3 GHz. We provide a detailed analysis of the frequency-dependent response within an equivalent circuit picture taking into account external contact contributions. In addition, we performed two crucial experiments to check for electrode contributions, namely measurements of a sample with varying contact material and with varying sample thickness.

Powders of CCTO were synthesized by standard solid state reaction, starting from well ground stoichiometric mixtures of $CaCO_3$, $CuO$, and $TiO_2$. Samples were heated in alumina crucibles at 1000 °C for 48 h with one intermediate grinding. Powder X-ray diffraction proved CCTO to be single phase. A structure analysis by Rietveld refinement yielded good agreement with previous reports based on neutron diffraction data [18]. For the dielectric measurements metallic contacts were applied to adjacent sides of the pellets forming a parallel-plate capacitor. Various types of contacts were used, namely polished brass plates pressed onto the sample, silver paint, and sputtered gold or silver (thickness 100 nm). The conductivity and permittivity were measured over a broad frequency range of more than 10 decades (0.1 Hz $< n <$ 1.3 GHz) at temperatures down to 20 K. A frequency response analyzer (Novocontrol α-analyzer) was used at $n <$ 1 MHz and a reflectometric technique, employing an impedance analyzer (Agilent E 4991A), at $n >$ 1 MHz [19].

Figure 1 shows the temperature dependence of the dielectric constant $e'$ and conductivity $s'$ for various frequencies for a sample with sputtered silver contacts. Similar to earlier reports, $e'(T)$ [Fig. 1(a)] exhibits a step like

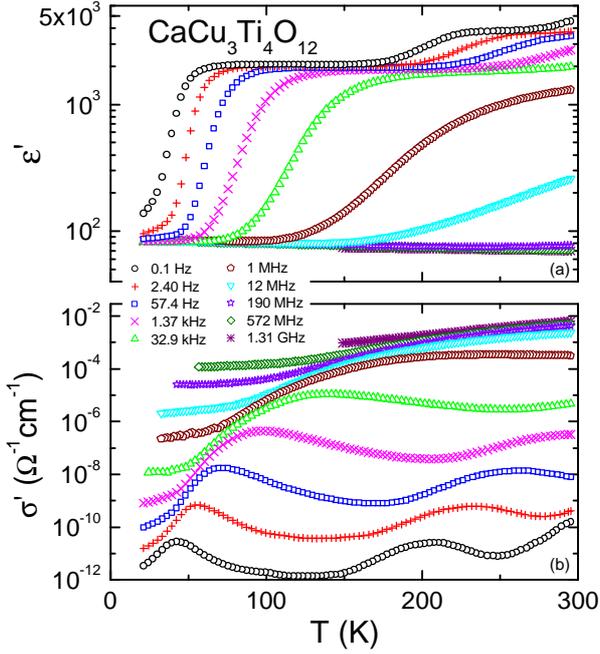

FIG. 1 (color online). Temperature dependence of dielectric constant (a) and conductivity (b) of CCTO for various frequencies (contacts: sputtered Ag).

increase from a low-temperature plateau value of the order of 100 towards "colossal" values of $\epsilon'$ at high temperatures. The step shifts to higher temperatures with increasing frequency and is accompanied by a peak in $\sigma'(T)$ [Fig. 1(b)] (as $\sigma' \sim \epsilon'' \times \omega$, this corresponds to a peak in $\epsilon''(T)$, too). Overall, $\epsilon'(T)$ and $\sigma'(T)$ show the typical signature of relaxational behavior as it is commonly observed e.g. for the glassy freezing of dipolar molecules [20]. Compared to the work of Homes *et al.* [3], two differences become obvious: i) In the present results instead of a single relaxation feature, a second step in $\epsilon'(T)$ occurs at high temperatures and correspondingly a second high-temperature peak in $\sigma'(T)$ shows up. ii) While the absolute values of the low-frequency plateau, $\epsilon'_{low} \approx 80$ are similar, the value of the high-frequency plateau, $\epsilon'_{high} \approx 4000$ is smaller. Concerning point i) it should be noted that in various other reports on the dielectric response of CCTO [2,7,8,11] two steps in $\epsilon'(T)$ [or correspondingly in $\epsilon'(\nu)$] were detected, too. Interestingly, all these reports were on polycrystalline samples as in the present work. Thus one can conclude that the second step in $\epsilon'$ is non-intrinsic and connected to the occurrence of grain boundaries in polycrystalline samples. Concerning point ii), we want to remark that there is a large variation in reported $\epsilon'_{high}$-values in literature with values varying between 500 and 80000 [1,2,3,5,6,7,8,11,12] and with our value lying well within this range. This broad variation is puzzling and a first hint that $\epsilon'_{high}$ may be due to non-intrinsic effects. Finally, we want to point out that our high-frequency measurements reveal CCTO having no colossal $\epsilon'$ at room temperature for GHz frequencies [Fig. 1(a)]. In addition, for $\nu \geq$ MHz we find $\epsilon'$ being strongly temperature dependent. Thus, even if the detected CDCs should be intrinsic, CCTO certainly is not suitable for applications in modern high-frequency electronics.

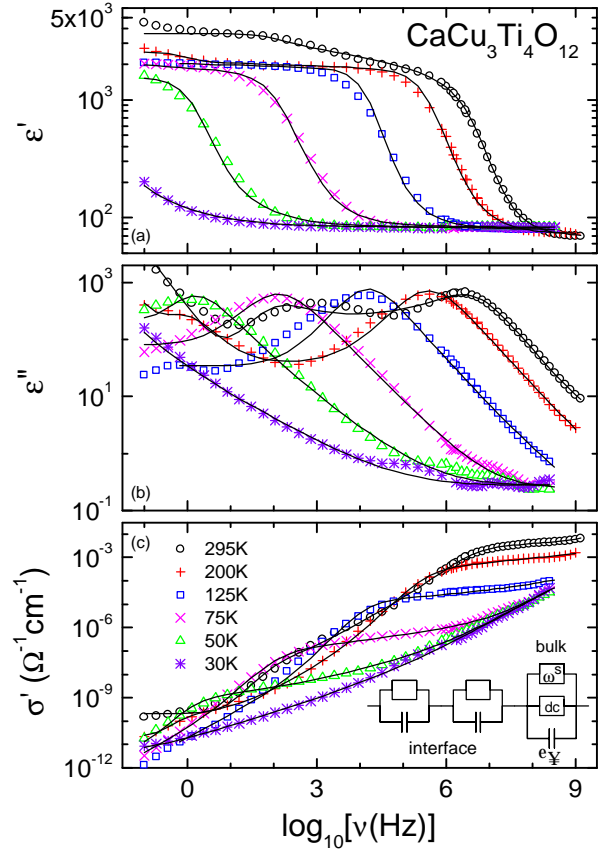

FIG. 2 (color online). Frequency dependence of dielectric constant (a), loss (b), and conductivity (c) of CCTO for selected temperatures (same measurement as Fig. 1). The lines are fits with the equivalent circuit indicated in frame (c).

In Fig. 2 the frequency dependence of the complex permittivity and the conductivity is shown. Again the results are qualitatively consistent with those reported in literature [2,3,5,8]. In the following we will promote the view that the main relaxational step in $\epsilon'$ is due to external contacts. As known since long, at the electrode/bulk interface thin depletion layers can arise, e.g. due to the formation of Schottky diodes [16,17]. These thin layers of low conductivity act as high capacitance in parallel with a large resistor, connected in series to the bulk sample, which can lead to the erroneous detection of very large values of $\epsilon'$ (for details, see [15]). Thus we employed the equivalent circuit shown in Fig. 2(c), to fit the data of Fig. 2. Here two parallel RC-circuits were used to account for the external and internal (grain boundary) contacts at $T \geq 200$ K, respectively one RC-circuit for the lower temperatures, which were connected in series to the bulk response. At high frequencies the contact resistors become successively shorted by the contact capacitors and the intrinsic response is detected. In Fig. 2(c) it becomes obvious that the intrinsic $\sigma'$ increases with frequency at low temperatures, a behavior which is often observed in transition metal oxides and indicative of hopping conduction of Anderson-localized charge carriers [17,21]. It can be parameterized by the so-called universal dielectric response, namely a power-law increase of $\sigma'(\nu)$ and the corresponding contribution in $\varepsilon'(\nu)$ [17,22] and is taken into account by an appropriate element in the equivalent circuit. The lines in Fig. 2 are fits with this circuit. A satisfactory agreement with the experimental data could be achieved in this way; especially the apparently colossal values of $\varepsilon'_{high}$ are explained with a reasonable contact capacitance in the order of several 100 pF. Via the time constant of the circuit, the semiconducting characteristic of the bulk conductivity leads to the observed strong temperature shift of the step. As contact steps and peaks are rather broad, a distribution for the time constant of the leaky capacitor representing the barrier had to be used, which can be ascribed to the roughness of the sample surface. Here for simplicity reasons a Cole-Cole distribution was employed [17]. This obviously is not quite correct as some deviations of fit and data especially at the left wing of the $\varepsilon''$-peaks occur [Fig. 2(b)], but the intention was only to demonstrate the principle feasibility of this equivalent circuit description for the explanation of the CDCs and the relaxation-like behavior of CCTO.

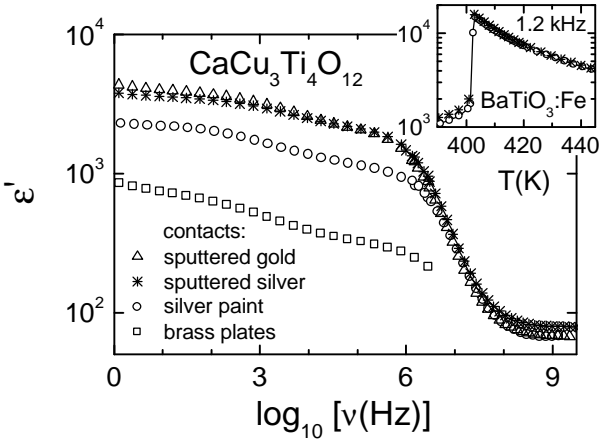

FIG. 3. Frequency-dependent dielectric constant at room temperature of a CCTO sample, successively prepared with different contact types. The inset shows $\varepsilon'(T)$ of Fe-doped ferroelectric BaTiO$_3$ close to $T_c$, measured with silver paint and sputtered silver contacts.

While this successful equivalent circuit description demonstrates that the picture of external contacts causing the CDCs of CCTO is in accord with the experimental data, it cannot be considered as a real proof of this notion. As noted before [15], only additional experiments, especially a variation of the contact type or of the sample thickness, can finally prove such a scenario. Thus we have investigated one sample with different types of contacts, namely i) polished brass plates pressed to the bare sample surface, ii) silver paint contacts, iii) sputtered silver contacts of 100 nm thickness, applied after removing the silver-paint in an ultrasonic bath, and finally, iv) sputtered gold contacts of the same thickness, applied after removing the silver by polishing. The results of the corresponding frequency sweeps, performed at room temperature, are shown in Fig. 3. Depending on the contact type, $\varepsilon'_{high}$ varies considerably, the contacts formed by brass plates leading to the lowest and the sputtered contacts to the highest values. This finding clearly proves that external contact contributions are causing the detected CDCs. In contrast to $\varepsilon'_{high}$, $\varepsilon'_{low}$ is nearly identical for the silver paint and sputtered contacts (for experimental reasons, brass plates could not be used at frequencies above MHz). Thus the value of $\varepsilon'_{low} \approx 80$ constitutes the intrinsic dielectric constant of CCTO. An $\varepsilon'_{low}$ of the order of 100 consistently was reported in numerous earlier reports on CCTO [2,3,5,7,8,11,12], in marked contrast to the mentioned strong variation of the reported magnitude of $\varepsilon'_{high}$. To preclude objections that silver paint contacts may not be suited to correctly measure high dielectric constants, the inset of Fig. 3 shows results on ferroelectric BaTiO$_3$ (slightly doped with Fe, which is not relevant here) obtained for the same sample with silver paint and sputtered silver contacts. The identical result of a truly intrinsic $\varepsilon'$ up to $10^4$ demonstrates the suitability of silver paint as contact material. The finding that sputtered contacts lead to the highest apparent dielectric constants can be explained by a more effective formation of the Schottky barriers because the very small metal clusters applied during sputtering will lead to a larger area of direct metal-semiconductor contact than for the relatively large particles ($\geq \mu m$) suspended in the silver paint. Further it is reasonable that metal plates simply pressed to the sample surface will lead to the poorest "wetting" and thus to the smallest values of $\varepsilon'$. The sputtered silver and gold contacts lead to almost identical results because the electron work functions of these metals, determining the thickness of the depletion layer of the Schottky diode, are almost identical.

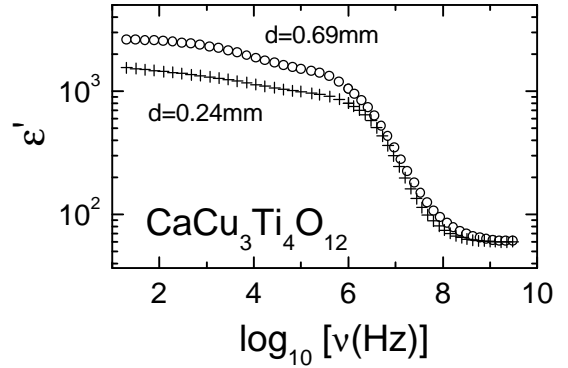

FIG. 4. Frequency-dependent dielectric constant at room temperature of a CCTO sample, measured before and after reducing its thickness by polishing. The contacts were formed by silver paint.

To further corroborate our findings, we also varied the thickness of one sample by polishing it down after measurement, and remeasuring it. Figure 4 shows the results: The value of $\varepsilon'_{high}$ is significantly higher for the thicker sample, which should not be the case for an intrinsic or IBLC origin of the CDCs. Instead this result can be understood assuming roughly the same depletion layer contribution to the measured capacitance in both cases. As $\varepsilon'$ is calculated from the capacitance by dividing it by the empty capacitance $C_0$, which is smaller for the thicker sample, of course a larger apparent dielectric constant results. The fact that the difference of $e'_{high}$ of both measurements comes out somewhat lower than expected from the thickness ratio may be due to a different roughness of the surface after polishing.

In summary, our results clearly prove that external contact contributions lead to the erroneous detection of CDCs in CCTO. This finding is in stark contrast to conclusions drawn in ref. [8], also based on a variation of contacts. In this work a thin insulating layer of $Al_2O_3$ was deposited onto the sample surface before applying metallic contacts. The idea was to exclude the formation of Schottky diodes by avoiding a direct metal-semiconductor contact. In this way, still rather high values of $e'_{high} \approx 5000$ were detected, for which an intrinsic nature was claimed. However, the arguments of the authors are only correct assuming two *loss-free* capacitors, namely bulk and layer in series arrangement, in which case the smaller bulk capacitance dominates. However, taking into account that the bulk material, CCTO, has to be presented by a *leaky* capacitor, leads to the same overall behavior of $C(n)$ as for Schottky-type contacts. Especially, highly enhanced apparent values of $e'$ are measured for $n \to 0$, due to the high capacitance of the very thin (100 nm) insulating layer. In this configuration, the apparent dielectric constant $e'_a$ is given by $e'_a = e'_l \times t_b/t_l$, with $e'_l$ the dielectric constant of the insulating layer, and $t_b$ and $t_l$ the thickness of bulk and layer, respectively. Taking $e'_l \approx 10$ ($Al_2O_3$), $t_l = 100$ nm and assuming $t_b = 0.1$ mm, a value of $e'_a \approx 10^4$ results, which is in the same order of magnitude as the values reported in [8]. In an earlier work [23] we already demonstrated that indeed even such an insulating layer (in our case Teflon or mica foils of several 10 μm on semiconducting $CdF_2$:In) can lead to very large apparent dielectric constants and a Maxwell-Wagner relaxation in the spectra. We want to point out that the view of external contacts causing the CDC in CCTO is in good accord also with the dielectric results of ref. [5], which were analyzed employing complex impedance plots. It is well known that semicircular arcs occurring in the complex impedance plane at low frequencies can be indicative of interfacial contributions, both from grain boundaries or external contacts. The authors of ref. [5] suggested a grain boundary origin but of course the observed features could be also due to external contacts. The dependence on annealing time and oxygen partial pressure during annealing found in [5] can be ascribed to a variation of external contacts, e.g. via a variation of oxygen stoichiometry at the sample surface.

In conclusion, we have performed two crucial experiments to clarify the true origin of the CDCs in CCTO. The dependence of $e'_{high}$ both on contact type and sample thickness clearly proves that external contacts via an interfacial polarization process lead to the detection of very high dielectric constants. The observed relaxation-like spectral features are of Maxwell-Wagner type and easily understandable within an equivalent circuit picture. Within this framework, it also becomes clear why the reported values of $e'_{high}$ vary so much: Its absolute value depends on details of contact formation, e.g. the surface roughness, type of contact, and stoichiometry at the surface. We also believe that CCTO is not exceptional in any way compared to other isostructural compounds. The differences in $e'$ at 100 kHz found for 21 related materials by Submaranian et al. [4] with reported values between 33 and 3,560, CCTO standing out with $e' = 10,286$, may also be explainable by different contact contributions. Also the frequency position of the main relaxational step may vary (via the circuits time constant, it depends also on the bulk conductivity) in a way that at 100 kHz sometimes $e'_{low}$ is measured or even a value in the transition region between $e'_{low}$ and $e'_{high}$. In addition, recently it was demonstrated by broadband dielectric measurements of $Cu_2Ta_4O_{12}$, being structurally related to CCTO, that apparent CDCs can occur also in this material [13]. As in $Cu_2Ta_4O_{12}$, the intrinsic $e'$ (= $e'_{low}$) of CCTO as determined from literature and from our measurements consistently seems to be of the order of 100. This is not colossal but nevertheless rather high, compared to most other technically relevant materials and thus CCTO or related compounds still may be of interest for possible applications in electronics. It certainly also would be interesting to check what may cause this rather high dielectric constant, e.g. ionic or electronic polarization processes.

We thank U. Pamukçi for performing the $BaTiO_3$ measurements. This work was supported by the Deutsche Forschungsgemeinschaft via the Sonderforschungsbereich 484 and partly by the BMBF via VDI/EKM, FKZ 13N6917.